# Invariant indices of polarimetric purity. Generalized indices of purity for *n*x*n* covariance matrices

**I. San José**
*Instituto Aragonés de Estadística*
*Camino de las Torres 53, 50071 Zaragoza, Spain*
isanjose@aragon.es

**J. J. Gil**
*ICE Universidad de Zaragoza*
*c/ Pedro Cerbuna 12, 50009 Zaragoza, Spain*
ppgil@unizar.es

## Abstract

A proper set of indices characterizing the polarimetric purity of light and material media is defined from the eigenvalues of the corresponding coherency matrix. A simple and generalizable relation of these indices with the current parameters characterizing the global purity is obtained. A general definition for systems characterized by $n \times n$ positive semidefinite Hermitian matrices is introduced in terms of the corresponding eigenvalues and diagonal Gell-Mann matrices. The set of $n-1$ indices of purity has a nested structure and provide complete information about the statistical purity of the system.

## Contents

## 1 Introduction

This paper deals with the mathematical description of the second order statistical purity of systems characterized by $n\times n$ positive semidefinite Hermitian matrices, as is the case of density matrices in quantum mechanics and coherency matrices in polarization optics.

An overall invariant dimensionless measure of the statistical purity is given by the "Degree of Purity". This quantity is related to the Von-Neumann entropy [1,2]. Nevertheless, a complete description of purity requires a set of $n-1$ invariant dimensionless "indices of purity". These parameters have been defined by us in previous works for $n=2,3,4$ under the scope of polarization optics [3,2]. Here we introduce a complete description of the statistical purity based on a new generalized definition of the indices of purity. The nested structure of the indices of purity is studied as well as their relations with the degree of purity.

For the sake of clarity, the first sections of this paper deal with the case of 2D, 3D and 4D polarization optics, so that the concepts are finally generalized to any system characterized by a positive semidefinite Hermitian matrix (hereafter called coherency matrix).

## 2. 2D polarimetric purity

In the second order optics approach, the polarimetric purity of a plane wave is characterized through a unique parameter, namely the corresponding degree of polarization *P*, which can be written in terms of the eigenvalues of the corresponding $2\times 2$ coherency matrix $\mathbf{\Phi}$ [4,5]

$$P=\frac{\lambda_0-\lambda_1}{\lambda_0+\lambda_1}=\frac{\lambda_0-\lambda_1}{\mathrm{tr}\mathbf{\Phi}}, \tag{1}$$

where the energy flux per unit area (usually called "intensity" in this context) is given by $I=\mathrm{tr}\mathbf{\Phi}$. Thus, the quantities $I$ and $P$ are invariant in the sense that they remain unchanged under unitary transformations of the coherency matrix [6] and, hence, they are invariant with respect to changes of the reference system *XY*.

In agreement with the statistical nature of $\mathbf{\Phi}$, which is a covariance matrix (and, hence, is a positive semidefinite Hermitian matrix), $\lambda_1,\lambda_2$ are non-negative. Pure states are characterized by rank-1 polarization matrices (only one nonzero eigenvalue, $P=1$), whereas rank-2 polarization matrices correspond to mixed states ($P<1$).

An appropriate expression of *P* that will be useful for future considerations is

$$P_{(2)}=\left(\frac{2\mathrm{tr}\left(\mathbf{\Phi}^2\right)}{\left(\mathrm{tr}\mathbf{\Phi}\right)^2}-1\right)^{1/2}, \tag{2}$$

where the subscript $_{(2)}$ has been added in order to compare these expressions with other that will appear concerning higher order coherency matrices.

Furthermore, Wolf, in a classic paper [7], showed that there always exist two orthogonal reference directions such that the degree of coherence $\mu$ reaches its maximum value, which coincides with $P$. Thus, $P$ can be also defined as the maximum modulus of the degree of coherence.

An alternative measure related with the purity of plane waves is given by the von Neumann entropy $S$ [8-11], which is defined as

$$S = -\frac{\mathrm{tr}\left(\mathbf{\Phi}\ln\mathbf{\Phi}\right)}{\mathrm{tr}\mathbf{\Phi}} = -\frac{\sum_{i=0}^{1}\left(\lambda_i \ln \lambda_i\right)}{\mathrm{tr}\mathbf{\Phi}}. \quad (3)$$

This dimensionless quantity is a measure of the difference in the amount of information between a pure state and a mixed state (both with the same intensity) and, hence, is closely related to $P$ by the expression

$$S = S_{(2)}\left(P\right) \equiv -\left\{\frac{1}{2}\left(1+P\right)\ln\left[\frac{1}{2}\left(1+P\right)\right] + \frac{1}{2}\left(1-P\right)\ln\left[\frac{1}{2}\left(1-P\right)\right]\right\}, \quad (4)$$

where the subscript $(2)$ has been placed in order to compare these expressions with other that will appear concerning entropies corresponding to higher order coherency matrices.

Therefore, $S_{(2)}$ is characterized univocally by $P$ and decreases monotonically as $P$ increases. The maximum $S_{(2)} = \ln 2$ corresponds to $P = 0$, whereas the minimum $S_{(2)} = 0$ is reached for $P = 1$, (i.e. when light is totally polarized, regardless of its spectral profile) [1].

## 3. 3D Polarimetric purity

The two-dimensional formalism considered in the previous section is valid when the propagation direction of light is fixed, which is the commonest physical situation of interest in polarimetry. In the most general case, the three components of the electric field vector of the light wave should be considered in order to describe the polarization state.

A first proper definition of the 3D degree of polarization was presented by Samson in 1973 [12] within the scope of geophysical studies of ultra-low frequency magnetic fields. This result was also obtained by Barakat by formulating the degree of polarization in terms of scalar invariants of the coherency matrix [9]. Nevertheless, the study of the 3D degree of polarization has recently attracted attention due to the advances in optical nanotechnologies and from the necessity of understanding polarization phenomena in fluctuating near fields and evanescent waves.

Under the second order optics approach, a complete formulation of the purity of the 3D states of polarization based on two relative differences of the eigenvalues of the coherency matrix (indices of purity) has been introduced by Gil, Correas, Melero and Ferreira [3]. More recently, Ellis, Dogariu, Ponomarenko and Wolf [13] have also presented two parameters based on two relative differences of the eigenvalues of the coherency matrix and, from this result, Hioe [14] has introduced a parameter called the "degree of isotropy". We refer also to the relevant approaches obtained by Réfrégier, Roche and Goudail [15] who have emphasized the necessity of three invariant quantities to characterize the "polarimetric contrast", so that different sets of three parameters defined from the 3D coherency matrix have been considered, as well as their relations with the different "degrees of polarization" defined by Barakat [9] and by Samson [12].

Another purity parameter has recently been introduced by Dennis [16] by means of averaging the 3D state of polarization due to a dipole over all scattering directions, which leads to a purity measure, which is not a unitary invariant of the coherency matrix.

Particularly relevant contributions concerning the concept of the 3D degree of polarization have been presented by Setälä, Shevchenko, Kaivola and Friberg [17,18] and Ellis and Dogariu [19].

Quantum and classical approaches to the 3D degree of polarization concept, defining it as a distance between distributions [20] and between correlation matrices have been presented by Luis [21,22].

### 3.1 3D Degree of purity

In second order optics, 3D polarization states are characterized by the corresponding "coherency matrix" or "polarization matrix" **R** and the 3D "degree of purity" [2] can be defined as [12,9,17]

$$P_{(3)} = \left| \left[ \frac{1}{2} \left( \frac{3\,\mathrm{tr}\left(\mathbf{R}^2\right)}{\left(\mathrm{tr}\mathbf{R}\right)^2} - 1 \right) \right]^{1/2} \right| \tag{5}$$

This invariant non-dimensional parameter is limited to the interval $0 \leq P_{(3)} \leq 1$. $P_{(3)} = 1$ corresponds to the case that $\mathbf{R}$ has only one nonzero eigenvalue (total polarimetric purity: the direction of propagation is constant and the electric field describes a well-defined polarization ellipse). $P_{(3)} = 0$ is reached when the three eigenvalues of $\mathbf{R}$ are equal (equiprobable mixture of states and zero correlation between the electric field components).

As in Ref [2] we prefer using the term "degree of purity" (which, in this case refers to polarimetric purity), rather than "degree of polarization", for $P_{(3)}$.

As Setälä et al. have pointed out [17,18], $P_{(3)}$ takes into account not only the purity of the mean polarization ellipse, but also the stability of the plane that contains the instantaneous components of the electric field of the wave. Thus, for unpolarized light whose propagation direction remains fixed, $P_{(2)} = 0$ whereas $P_{(3)} = 1/2$. It is clear that in the 3D description of polarization, new relevant quantities and peculiar properties arise that do not exist in the 2D model. Therefore, the existence of three eigenvalues leads to the fact that, unlike the 2D model, the overall degree of purity $P_{(3)}$ does not provide complete information of the polarimetric purity properties.

### 3.2 3D Indices of purity

In order to find adequate invariant non-dimensional quantities that contain a complete description of the purity of 3D polarization states, we return to $P_{(2)}$ and observe that it can be defined as a relative difference between the two eigenvalues, so that $0 \leq P_{(2)} \leq 1$.

In the light of the structure of the algebraic expressions of the eigenvalues of **R**, and by inspecting the various relative differences between them, we see that a convenient pair of "indices of purity" is defined as [3,2]

$$P_1 = \frac{\lambda_0 - \lambda_1}{\mathrm{tr}\mathbf{R}}, \quad P_2 = \frac{\lambda_0 + \lambda_1 - 2\lambda_2}{\mathrm{tr}\mathbf{R}}. \tag{6}$$

These non-dimensional quantities are restricted by the following limits

$$0 \le P_1 \le P_2 \le 1. \tag{7}$$

From the above equations, the following quadratic relation between $P_{(3)}$ and the two indices of purity $P_1, P_2$ is derived [2]

$$P_{(3)}^2 = \frac{1}{4}\left(3P_1^2 + P_2^2\right). \tag{8}$$

Another interesting expression of $P_{(3)}$, as a homogeneous quadratic measure of all the relative differences between the eigenvalues, is the following [2]

$$P_{(3)}^2 = \frac{1}{2}\sum_{\substack{i,j=0\\ i<j}}^{2} p_{ij}^2, \quad p_{ij} \equiv \frac{\lambda_i - \lambda_j}{\mathrm{tr}\mathbf{R}}. \tag{9}$$

Thus, the two indices of purity provide complete information about the polarimetric purity of the corresponding polarization state. The physically feasible region in the *purity space* $P_1, P_2$ is shown in Fig.1.

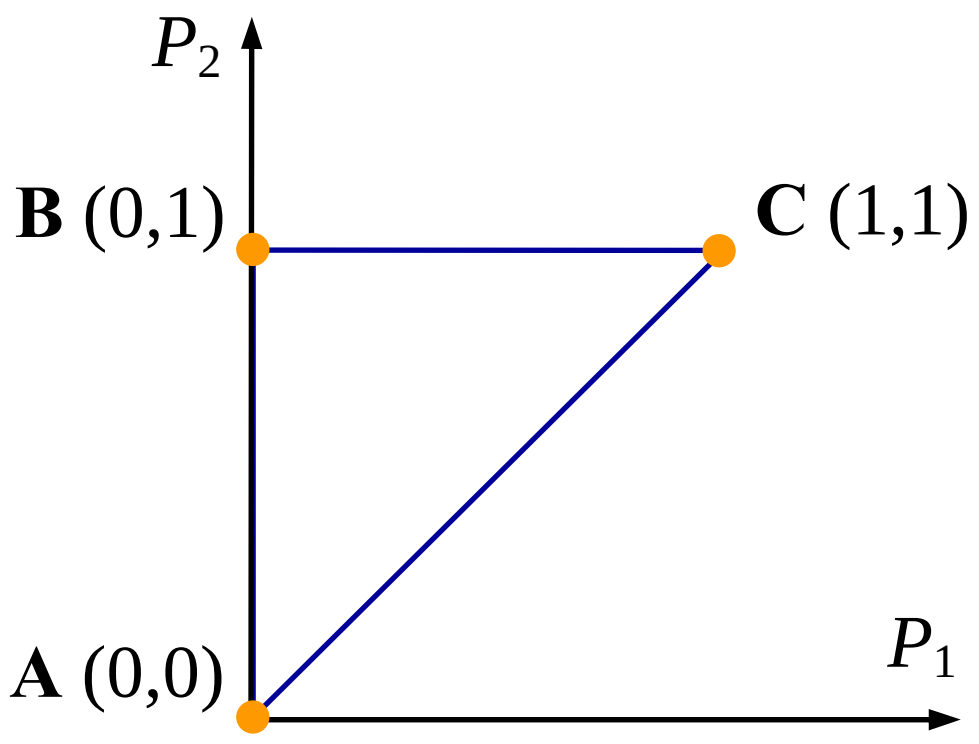

Fig.1. Feasible region for $P_1, P_2$ in the purity space.

As we will see in the case analysis presented below, $P_1$ maintain its meaning of “2D degree of polarization” whereas $P_2$ have the following properties [2]:

- If $P_1 = 0$ (fully random polarization ellipse), then $0 \le P_2 \le 1$. We see that the only possible contribution to purity of $P_2$ is related with the stability of the propagation direction.
- If $P_1 = 1$ (pure 2D state), then $P_2 = 1$ and $P_{(3)} = 1$ (3D pure state). This agrees with the fact that a 2D pure state is a 3D pure state with fixed direction of propagation.
- If $P_2 = 0$, then $P_1 = 0$ and $P_{(3)} = 0$ (3D unpolarized state). This agrees with the fact that a random propagation direction entails a random polarization ellipse.

These arguments indicate that $P_2$ is a measure of the degree of stability of the propagation direction of the wave.

A systematic analysis of the physical interpretation of $P_1, P_2$ for different cases has been performed by us in a previous paper [2] where the following names for the parameters of purity have been proposed, $P_1$: *degree of polarizat*ion; $P_2$: *degree of directionality,* and $P_{(3)}$: *degree of purity*. These names are consistent with the concepts underlying these relevant quantities. We also observe that these arguments agree with the conclusions of the study of Ellis and Dogariu concerning the concept of degree of polarization and the other invariant quantities describing the purity of a polarization state [19].

### 3.3 Von-Neumann entropy associated with R

From the description of the von Neumann entropy for *n*-dimensional density matrices introduced by Fano [8] and studied by Brosseau [1], the 3D polarization entropy $S$ can be defined as

$$S_{(3)} = -\frac{\mathrm{tr}\left(\mathbf{R}\ln\mathbf{R}\right)}{\mathrm{tr}\mathbf{R}} = -\frac{\sum_{i=0}^{2}\left(\lambda_i \ln \lambda_i\right)}{\mathrm{tr}\mathbf{R}}. \tag{10}$$

Thus, the entropy of a state of polarization can be expressed in terms of the indices of purity of **R** [2]

$$S_{(3)} = \\ = -\left\{\frac{\left(1+\frac{1}{2}P_2+\frac{3}{2}P_1\right)}{3}\ln\left[\frac{1}{3}\left(1+\frac{1}{2}P_2+\frac{3}{2}P_1\right)\right]+\frac{\left(1+\frac{1}{2}P_2-\frac{3}{2}P_1\right)}{3}\ln\left[\frac{1}{3}\left(1+\frac{1}{2}P_2-\frac{3}{2}P_1\right)\right]+\frac{\left(1-P_2\right)}{3}\ln\left[\frac{1}{3}\left(1-P_2\right)\right]\right\} \tag{11}$$

$S$ is characterized univocally by $P_1$ and $P_2$. The maximum $S_{(3)} = \ln 3$ corresponds to $P_1 = P_1 = P_{(3)} = 0$, whereas the minimum $S_{(3)} = 0$ is reached when $P_1 = P_1 = P_{(3)} = 1$.

Taking into account the expression of the von Neumann entropy associated with $2\times 2$ coherency matrices given by Eq. (4), and considering that, in the case of $3\times 3$ coherency matrices, there exist two indices of purity defined as relative differences of the eigenvalues, definitions for two respective "partial entropies", namely, *directional entropy* $S_{(2)}\left(P_2\right)$ and *polarization entropy* $S_{(2)}\left(P_1\right)$ have been introduced by us previously [2].

These invariant quantities contain objective information about the randomness in the propagation direction and in the polarization ellipse respectively. The condition $0 \le P_1 \le P_2 \le 1$ on the indices of purity has its counterpart in the partial entropies $0 \le S_{(2)}\left(P_2\right) \le S_{(2)}\left(P_1\right) \le \ln 2$.

Moreover, specific partial entropy can be defined for each purity parameter $p_{ij} = \dfrac{\lambda_i - \lambda_j}{\mathrm{tr}\,\mathbf{R}}$. Nevertheless, as we have seen, $P_1, P_2$ contain all the relevant information and, hence, $S_{(2)}(P_2)$, $S_{(2)}(P_1)$ are an adequate pair of representative quantities among all possible partial entropies $S_{(2)}(p_{ij})$.

## 4. Polarimetric purity of material media

The polarimetric properties of a linear passive medium are given by its corresponding Mueller matrix **M**. The information contained in **M** can also be represented by means of a positive semidefinite Hermitian matrix **H**, whose elements are defined through linear combinations of the elements of **M** [23].

### 4.1 Degree of polarimetric purity of material media

The degree of polarimetric purity $P_{(4)}$ of a material medium can be defined in terms of the coherency matrix as follows [23,2]

$$P_{(4)} = \left| \frac{1}{3} \left( \frac{4\mathrm{tr}(\mathbf{H}^2)}{(\mathrm{tr}\mathbf{H})^2} - 1 \right)^{1/2} \right| , \tag{12}$$

This invariant non-dimensional parameter is restricted to the interval $0 \le P_{(4)} \le 1$. The minimum $P_{(4)} = 0$ corresponds to an ideal total depolarizer, characterized by the fact that all eigenvalues of **H** are equal, i.e. the medium is composed of an equiprobable mixture of elements, and does not exhibit any polarimetric preference ($m_{ij} = 0$ except $m_{00}$). The maximum corresponds to a pure system ($\lambda_0 > 0,\ \lambda_1 = \lambda_2 = \lambda_3 = 0$).

$P_{(4)}$ gives an objective measure of the global polarimetric purity of the system, as well as of its depolarizing power, and provides criteria for the analysis of measured Mueller matrices [24-27].

Moreover, the "depolarizance" or "depolarizing power" $D$ of an optical system can be defined as [28,2]

$$D \equiv 1 - P_{(4)} . \tag{13}$$

### 4.2 Indices of polarimetric purity of material media

Given the statistical nature of the coherency matrices **H** representing material media, we emphasize the importance of obtaining parameters that give a measurement of their polarimetric purity. Usually the degree of purity $P_{(4)}$ [28,29] or, alternatively, the polarization entropy [30], is used as a quantity characterizing the overall purity.

Thus, in a similar manner that in the cases of $2\times2$ and $3\times3$ coherency matrices representing states of light, a complete description of the purity of **H** requires considering several relative differences between the four eigenvalues of **H**. Thus, three new invariant and non-dimensional "indices of purity" can be defined from the eigenvalues of **H** so that this set of three quantities contains all the information concerning the polarimetric purity. It should be noted that neither $P_{(4)}$ nor the polarization entropy cover all the information mentioned, but they can be calculated from the indices of purity.

In previous papers, we defined the following set of 4D indices of purity [3, 2]

$$p_1 \equiv \frac{\lambda_0 - \lambda_1}{\mathrm{tr}\mathbf{H}} \ , \ p_2 \equiv \frac{(\lambda_0 + \lambda_1) - (\lambda_2 + \lambda_3)}{\mathrm{tr}\mathbf{H}} \ , \ p_3 \equiv \frac{\lambda_2 - \lambda_3}{\mathrm{tr}\mathbf{H}} . \tag{14}$$

An exhaustive analysis of the properties of these parameters was presented in Ref [2]. The 3D model is reproduced when $\lambda_2 = \lambda_3$ or, equivalently, $p_3 = 0$. Pure systems are characterized by $P_{(4)} = p_1 = p_2 = 1, \ p_3 = 0$. Moreover, the values $P_{(4)} = p_1 = p_2 = p_3 = 0$ correspond to certain equiprobable mixtures of four (or more) incoherent elements, resulting in a Mueller matrix **O** whose elements are zero except for $o_{00}$. We observe that the behavior of $p_3$ is different from that of $p_1, p_2$ and corresponds to certain kind of "residual purity" [2]. A later study of these indices has lead us to the following more convenient definition

$$P_1 \equiv \frac{\lambda_0 - \lambda_1}{\mathrm{tr}\mathbf{H}} \ , \ P_2 \equiv \frac{\lambda_0 + \lambda_1 - 2\lambda_2}{\mathrm{tr}\mathbf{H}} \ , \ P_3 \equiv \frac{\lambda_0 + \lambda_1 + \lambda_2 - 3\lambda_3}{\mathrm{tr}\mathbf{H}} , \tag{15}$$

so that the eigenvalues can be expressed as

$$\begin{aligned}
\lambda_0 &= \frac{1}{4}\mathrm{tr}\mathbf{H}\left(1 + 2P_1 + \frac{2}{3}P_2 + \frac{1}{3}P_3\right), \\
\lambda_1 &= \frac{1}{4}\mathrm{tr}\mathbf{H}\left(1 - 2P_1 + \frac{2}{3}P_2 + \frac{1}{3}P_3\right), \\
\lambda_2 &= \frac{1}{4}\mathrm{tr}\mathbf{H}\left(1 - \frac{4}{3}P_2 + \frac{1}{3}P_3\right), \\
\lambda_3 &= \frac{1}{4}\mathrm{tr}\mathbf{H}\left(1 - P_3\right).
\end{aligned} \tag{16}$$

The indices of purity can be interpreted as probabilistic relative measures, which provide complete information about the relative amounts of the "equivalent pure components" of the target [2].

From the above equations, the following quadratic relation between $P_{(4)}$ and the three indices of purity $P_1, P_2, P_3$ is obtained

$$P_{(4)}^2 = \frac{1}{3}\left(2P_1^2 + \frac{2}{3}P_2^2 + \frac{1}{3}P_3^2\right). \tag{17}$$

Another interesting expression of $P_{(4)}$ as a symmetric quadratic mean of all the relative differences between pairs of eigenvalues is given by [2]

$$P_{(4)}^2 = \frac{1}{3}\sum_{\substack{i,j=0 \\ i<j}}^{3} p_{ij}^2 , \quad p_{ij} \equiv \frac{\lambda_i - \lambda_j}{\mathrm{tr}\mathbf{H}} . \tag{18}$$

Pure systems are characterized by $P_{(4)} = P_1 = P_2 = P_3 = 1,$. On the other hand, the values $P_{(4)} = P_1 = P_2 = P_3 = 0$ correspond to certain equiprobable mixtures of four (or more) incoherent elements, resulting in a Mueller matrix **O** whose elements are zero except for $o_{00}$.

By applying the starting conditions for the eigenvalues $0 \le \lambda_3 \le \lambda_2 \le \lambda_1 \le \lambda_0$, we find that the indices of purity are restricted by the following conditions

$$0 \le P_1 \le P_2 \le P_3 \le 1 . \quad (19)$$

Thus, if $P_1 = 1$, then $P_2 = P_3 = P_{(4)} = 1$. Moreover, if $P_3 = 0$, then $P_1 = P_2 = P_{(4)} = 0$. Total purity is characterized by $P_1 = 1 \left( P_{(4)} = 1 \right)$ whereas an equiprobable mixture of states is characterized by $P_3 = 0 \left( P_{(4)} = 0 \right)$.

### 4.3 4D purity space

Fig.2 shows the feasible region of the purity indices in the *purity space*. The restriction to the plane $P_3 = 1$ reproduces the feasible region for the indices of purity $P_1, P_2$ corresponding to $3\times3$ coherency matrices, and the feasible region for the degree of polarization of $2\times2$ coherency matrices corresponds to the segment BC $\left( P_2 = 1,\ 0 \le P_1 \le 1 \right)$.

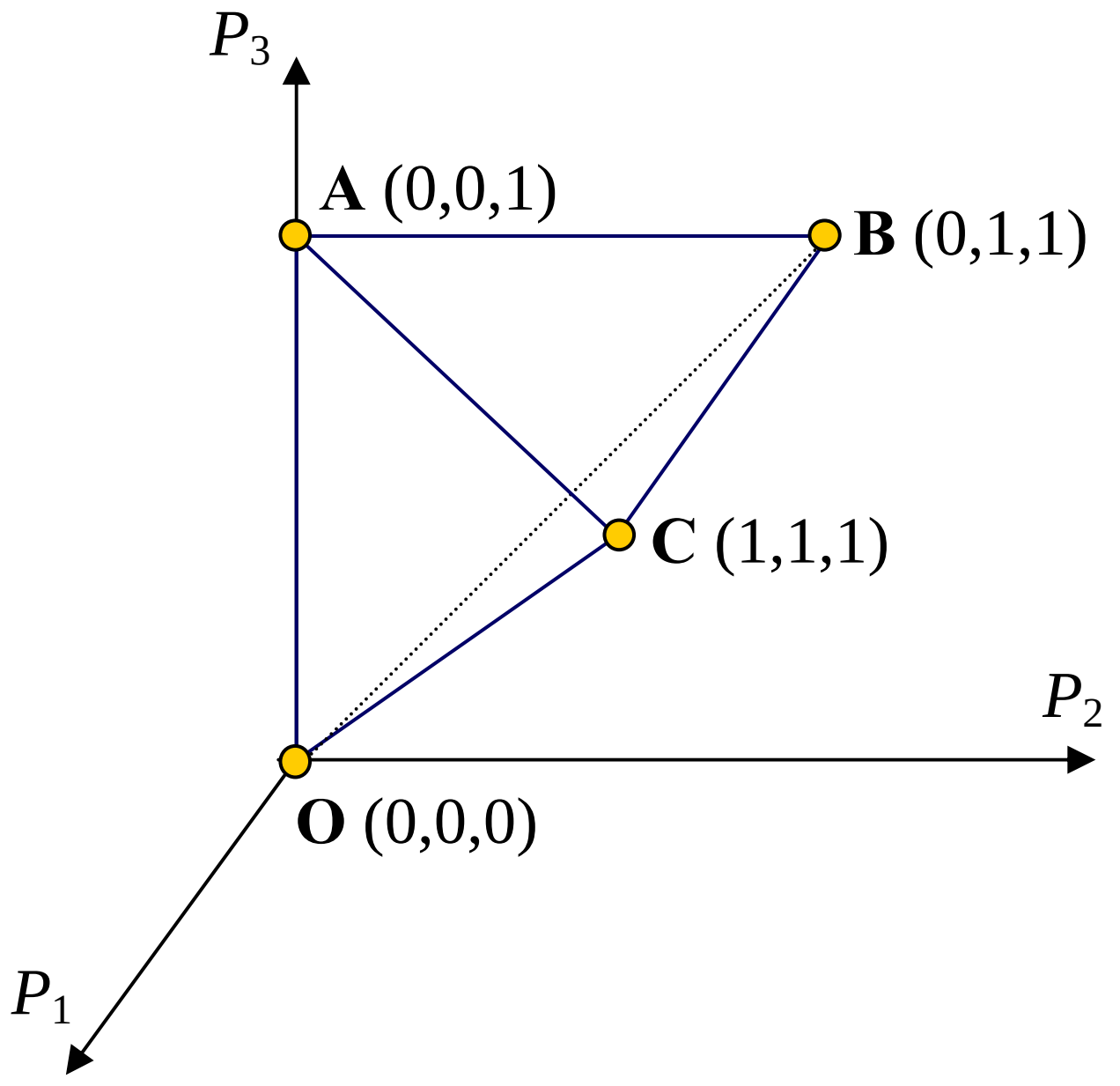

Fig.2. Feasible region for $P_1, P_2, P_3$ in the purity space.

Fig.2 summarizes the different physically realizable possibilities in terms of the values of the indices of purity. Nevertheless, it is worth considering some particular cases in order to understand the physical meaning of $P_i$, as well as to reproduce the feasible regions for 3D and 2D representations of the indices of purity.

*a)* The face CBA corresponds to states with $P_3 = 1$. Here we analyze the ranges $0 \le P_1 \le P_2 \le 1 \quad \left( 0 < P_{(4)} \le 1 \right)$.

The system is composed of one, two or three pure elements. The feasible region is determined by the points placed on the triangle CBA. This case is mathematically equivalent to that obtained for $3\times 3$ coherency matrices (3D polarization states).

*a.1)* $P_3=1,\ 0<P_2<1,\ P_1=0,\quad \left(1/3<P_{(4)}<1/\sqrt{3}\right)$.

The system is composed of two or three pure elements. The two more significant have equal cross sections $\lambda_0=\lambda_1$. The feasible region is the edge AB (vertices A and B excluded). These states are mathematically equivalent to states of light with random ellipse and whose direction of propagation is not constant.

*a.2)* $P_3=1,\ P_2=1,\ 0\le P_1\le 1\quad \left(1/\sqrt{3}\le P_{(4)}\le 1\right)$.

The system is composed of two pure elements. The feasible region is the edge BC and $P_1$ is mathematically equivalent to the 2D degree of polarization $P_{(2)}$.

*a.2.1)* $P_3=1,\ P_2=P_1=1\quad \left(P_{(4)}=1\right)$.

Pure system (non-depolarizing deterministic system) characterized by a Mueller-Jones matrix. The vertex C represents this state.

*a.2.2)* $P_3=1,\ P_2=1,\ P_1=0\quad \left(P_{(4)}=1/\sqrt{3}\right)$.

The system is composed of two pure elements with equal cross sections. The point B represents this state and the case is mathematically equivalent to 2D unpolarized light.

*b)* $P_1=0$. Here we consider the ranges $0<P_2\le P_3<1,\ \left(0<P_{(4)}<1/\sqrt{3}\right)$

The system is composed of four pure elements. The two more significant have equal cross sections $\lambda_0=\lambda_1$. The face OBA (edges OA and AB excluded) determines the feasible region. These states are exclusive of the 4D representation because the reproduction of 3D states (and, hence, possible 2D states) requires $\lambda_3=0\ \left(P_3=1\right)$.

*b.1)* $P_1=0,\ 0<P_2=P_3<1\quad \left(P_{(4)}<1/\sqrt{3}\right)$.

The system is composed of four pure elements where the two more significant have equal cross sections ($\lambda_0=\lambda_1$) and the two less significant have equal cross sections ($\lambda_2=\lambda_3$) with $\lambda_0\ne\lambda_2$. The feasible region is represented by the edge OB (vertices O, B excluded).

*c)* The face OCA (edges CA and OA excluded) corresponds to states with $0<P_1=P_2\le P_3<1\quad \left(0\le P_{(4)}<1\right)$.

The system is composed of four elements. The second and the third less significant have equal cross sections ($\lambda_1=\lambda_2$). These states are exclusive of the 4D representation.

*c.1)* $0<P_1=P_2=P_3<1\quad \left(0<P_{(4)}<1\right)$.

The feasible region corresponds to the edge OC (vertices O, C excluded) and represents states where the three less significant pure components have equal cross sections ($\lambda_1=\lambda_2=\lambda_3$) but different than $\lambda_0$.

*d)* The face OCB (edges OC and CA excluded) corresponds to states with

$0 \le P_1 < P_2 = P_3 < 1 \quad \left(0 < P_{(4)} < 1\right).$

The system is composed of four elements. The two less significant have equal cross ($\lambda_2 = \lambda_3$). These states are exclusive of the 4D representation.

*e)* $P_1 = P_2 = 0,\; 0 < P_3 < 1 \quad \left(0 < P_{(4)} < 1/3\right).$

These states are represented by the edge OA (vertices excluded) and correspond to an equivalent system composed of four pure elements, where the three more significant have equal cross sections ($\lambda_0 = \lambda_1 = \lambda_2$) but different than $\lambda_3$.

*f)* $P_1 = P_2 = P_3 = 0 \quad \left(P_{(4)} = 0\right).$

This system, equivalent to an ideal depolarizer, is composed of four pure elements with equal cross sections and is represented by the vertex A. All the elements of the Mueller matrix are zero except for $m_{00}$.

### 4.4 Polarization entropy

Polarization entropy [30] is a concept related with non-purity of the material samples [2]. This non-dimensional parameter, which is directly related with the degree of purity, is defined as

$$S_{(4)}(\mathbf{H}) = -\frac{\mathrm{tr}\left(\mathbf{H}\ln\mathbf{H}\right)}{\mathrm{tr}\mathbf{H}} = -\frac{\sum_{i=0}^{3}\left(\lambda_i \ln \lambda_i\right)}{\mathrm{tr}\mathbf{H}}. \qquad (20)$$

The indices of purity provide complete information about the depolarization properties and about the polarimetric purity of material samples, in such a manner that they cover a scope of information wider than $S_{(4)}$ and $P_{(4)}$. In fact, $S_{(4)}$ and $P_{(4)}$ can be directly obtained from the indices of purity.

Following the idea introduced in the section devoted to the entropy of $3\times 3$ coherency matrices, it is possible to extend to $4\times 4$ coherency matrices the concept of partial entropies by defining the following quantities: *polarization entropy* $S_{(2)}\left(P_1\right)$, 3D *directional entropy* $S_{(2)}\left(P_2\right)$, and 4D *directional entropy* $S_{(2)}\left(P_3\right)$. The latter is exclusive of $n\times n$ coherency matrices with $n \ge 4$.

## 5 *n*D Polarization algebra

Through the previous sections, we have seen that there exists a strong and clear symmetry in the description of the purity properties [2] in polarization optics. The expansion of the $n\times n\ \left(n>1\right)$ coherency matrix in a basis composed the $n\times n$ identity matrix together with a set of $n-1$ trace orthogonal, traceless, Hermitian matrices (generalized Gell-Mann matrices) leads to the corresponding *n*D Stokes parameters. The overall polarimetric purity is defined by means of the degree of purity $P_{(n)}$, whereas the detailed information about purity is given by the corresponding $\left(n-1\right)$ set of indices of purity [2].

Thus, if $\mathbf{\Omega}$ is a $nxn$ positive semidefinite Hermitian matrix and $\mathbf{\Theta}_{ij}\ \left(i,j=0,n-1\right)$ is a basis of Hermitian, trace orthogonal, $nxn$ matrices composed of the $n-1$ traceless generators of the SU($n$) group plus the identity matrix, all of them satisfying $\mathbf{\Theta}_{ij}^2=\mathbf{1}$ and $\operatorname{tr}\left(\mathbf{\Theta}_{ij}\mathbf{\Theta}_{kl}\right)=n\delta_{ik}\delta_{jl}$, the real coefficients $c_{ij}=\operatorname{tr}(\mathbf{\Theta}_{ij}\mathbf{\Omega})$ of the expansion

$$\mathbf{\Omega}=\frac{1}{n}\sum_{i,j=0}^{n-1}c_{ij}\mathbf{\Theta}_{ij}\ , \tag{21}$$

are the measurable quantities characterizing completely the second order properties of the system.

A proper choice for $\mathbf{\Theta}_{ij}$ is the set constituted by the $n\times n$ identity matrix together with the set of $n-1$ generalized Gell-Mann matrices [31], adequately weighted in order to ensure the normalization condition $\operatorname{tr}\left(\mathbf{\Theta}_{ij}\mathbf{\Theta}_{kl}\right)=n\delta_{ik}\delta_{jl}$. As an important observation, it should be noted that, in the $4\times4$ case, the choice of a non-Gell-Mann basis composed of the set of "modified Dirac matrices" [32,2]

$$\mathbf{E}_{ij}=\boldsymbol{\sigma}_i\otimes\boldsymbol{\sigma}_j\ \left(i,j=0,1,2,3\right), \tag{22}$$

given by Kronecker products of the $\boldsymbol{\sigma}_i$ matrices

$$\boldsymbol{\sigma}_0=\begin{pmatrix}1&0\\0&1\end{pmatrix},\quad \boldsymbol{\sigma}_1=\begin{pmatrix}1&0\\0&-1\end{pmatrix},\quad \boldsymbol{\sigma}_2=\begin{pmatrix}0&1\\1&0\end{pmatrix},\quad \boldsymbol{\sigma}_3=\begin{pmatrix}0&-i\\i&0\end{pmatrix}, \tag{23}$$

where $\boldsymbol{\sigma}_1,\boldsymbol{\sigma}_2,\boldsymbol{\sigma}_3$ are the Pauli matrices (that is to say, the $2\times2$ Gell-Mann matrices), leads to the following expansion of $\mathbf{\Omega}$

$$\mathbf{\Omega}=\frac{1}{4}\sum_{i,j=0}^{3}m_{ij}\mathbf{E}_{ij}\ ,\ m_{ij}=\operatorname{tr}(\mathbf{E}_{ij}\mathbf{H}), \tag{24}$$

where $m_{ij}$ are the sixteen real elements of the corresponding Mueller matrix. It should be noted that, whereas the basis $\mathbf{\Theta}_{ij}$ can be generalized to $n$D, this is not possible for the basis $\mathbf{E}_{ij}$, which only appear as peculiarity of 4D.

In order to clarify this discussion, in the Appendix we present explicit expressions of both sets of matrices.

### 5.1 Generalized degree of purity

The degree of purity is defined as the following invariant non-dimensional quantity [2]

$$P_{(n)}=\left\{\frac{1}{n-1}\left[\frac{n\left[\operatorname{tr}\left(\mathbf{\Omega}^2\right)\right]}{(\operatorname{tr}\mathbf{\Omega})^2}-1\right]\right\}^{\frac{1}{2}}, \tag{25}$$

which can also be expressed as

$$P_{(n)} = \left| \left[ \frac{1}{n-1} \left( \frac{n\|\mathbf{\Omega}\|_2^2}{\|\mathbf{\Omega}\|_0^2} - 1 \right) \right]^{1/2} \right|, \tag{26}$$

in terms of the two following norms of $\mathbf{\Omega}$ [2]

$$\|\mathbf{\Omega}\|_2 \equiv \left| \left[ \mathrm{tr}\left(\mathbf{\Omega}^2\right) \right]^{1/2} \right|, \quad \|\mathbf{\Omega}\|_0 \equiv \mathrm{tr}\mathbf{\Omega} = \left\| \sqrt{\mathbf{\Omega}} \right\|_2^2. \tag{27}$$

$P_{(n)}$ gives a global measure of the purity of the system, so that $0 \le P_{(n)} \le 1$. The minimum value of the degree of purity $P_{(n)} = 0$ corresponds to a fully random system, whereas the maximum value $P_{(n)} = 1$ corresponds to a pure system.

The degree of purity can also be expressed as

$$P_{(n)}^2 = \frac{1}{n-1} \sum_{\substack{i,j=0 \\ i<j}}^{n} p_{ij}^2, \quad p_{ij} \equiv \frac{\lambda_i - \lambda_j}{\mathrm{tr}\mathbf{\Omega}}. \tag{28}$$

## 5.2 Generalized indices of purity

The indices of purity defined in previous sections under the scope of optical polarimetry can be formulated in a general way for systems characterized by $n \times n$ covariance matrices. This general definition, which is presented in this section, is the main result of this paper.

Given a positive semidefinite, $n \times n \left(n > 1\right)$, Hermitian matrix $\mathbf{\Omega}$ and $\mathbf{\Lambda} \equiv \mathrm{diag}\left(\lambda_0, \lambda_1, ..., \lambda_{n-1}\right)$ being the corresponding diagonal $n \times n$ matrix whose nonzero elements are the ordered eigenvalues $\lambda_0 \ge \lambda_1 \ge .... \lambda_{n-1} \ge 0$ of $\mathbf{\Omega}$, we define the following "indices of purity" (IoP)

$$P_k = \sqrt{\frac{k^2+k}{n}} \left[ \frac{\mathrm{tr}\left(\mathbf{\Lambda}\,\mathbf{\Theta}_{kk}\right)}{\mathrm{tr}\left(\mathbf{\Omega}\right)} \right], \quad k = 1, ..., n-1, \tag{29}$$

where $\mathbf{\Theta}_{kk}$ are the $\left(k+1\right)$-rank, diagonal, Gell-Man matrices corresponding to the representation of the diagonal generators of the group $\mathrm{SU}\left(k+1\right)$, in the same way as they are described in [33] and in [31], but adequately normalized so that $\mathrm{tr}\left(\mathbf{\Theta}_{kk}^2\right) = n$,

$$\mathbf{\Theta}_{kk} = \sqrt{\frac{n}{k^2+k}} \begin{bmatrix} 1 & & & & & & \\ & \cdot & & & & & \\ & & 1 & & & & \\ & & & -k & & & \\ & & & & 0 & & \\ & & & & & \cdot & \\ & & & & & & 0 \end{bmatrix}. \tag{30}$$

Thus, each index $P_k$ can be written as

$$P_k = \frac{\sum_{i=0}^{k-1} \lambda_i - k\,\lambda_k}{\mathrm{tr}(\mathbf{\Omega})} \quad k = 1,\ldots,n-1. \tag{31}$$

These indices are restricted by the following nested limits

$$0 \le P_1 \le \ldots \le P_{n-1} \le 1. \tag{32}$$

We see that, if $P_1 = 1$, then $P_1 = \ldots = P_{n-1} = 1$, and if $P_{n-1} = 0$, then $P_1 = \ldots = P_{n-1} = 0$.

In order to construct a general expression of the eigenvalues of $\mathbf{\Omega}$ in function of these indices, for $n > 1$ we define the parameter $P_0 = 0$, so that

$$\begin{aligned} \lambda_k &= \mathrm{tr}(\mathbf{\Omega})\left[\frac{1}{n} - \frac{P_k}{k+1} + \sum_{i=k+1}^{n-1} \frac{P_i}{i(i+1)}\right] \quad (0 \le k < n-1), \\ \lambda_{n-1} &= \frac{\mathrm{tr}(\mathbf{\Omega})}{n}\left[1 - P_{n-1}\right]. \end{aligned} \tag{33}$$

For $n > 1$, we also define the additional operational parameters $P_n = 1$ and $\lambda_n = 0$, so that the difference between two consecutive indices can be expressed as

$$P_{k+1} - P_k = \frac{(k+1)}{\mathrm{tr}(\mathbf{\Omega})}\left(\lambda_k - \lambda_{k+1}\right), \; 0 \le k \le n-1. \tag{34}$$

We also observe the following convexity relation of these differences

$$\sum_{k=0}^{n-1}\left(P_{k+1} - P_k\right) = 1. \tag{35}$$

The degree of purity can be written in function of the IoP as follows

$$P_{(n)} = \left|\left\{\frac{n}{n-1}\left[\sum_{k=1}^{n-1} \frac{P_k^2}{k(k+1)}\right]\right\}^{1/2}\right|. \tag{36}$$

From this expression, it is straightforward to obtain the following conditions

$$\begin{aligned} P_{(n)} &= 0 \Leftrightarrow P_1 = P_2 = \ldots = P_{n-1} = 0, \\ P_{(n)} &= 1 \Leftrightarrow P_1 = P_2 = \ldots = P_{n-1} = 1. \end{aligned} \tag{37}$$

Hence, total purity implies that all partial purities are equal to one, and null purity implies that all partial purities are equal to zero.

Given the nested structure of the purity algebra, it is interesting to observe that the *n*-dimensional degree of purity can be expressed in terms of the (*n*-1)-dimensional degree of purity and the index of purity $P_{n-1}$

$$P_{(n)}^2 = \frac{n(n-2)}{(n-1)^2} P_{(n-1)}^2 + \frac{1}{(n-1)^2} P_{n-1}^2. \tag{38}$$

Since the Hermitian matrix $\mathbf{\Omega}$ is positive semidefinite, it can be diagonalized through a unitary transformation and can be written as $\mathbf{\Omega} = \mathbf{U}\mathbf{\Lambda}\mathbf{U}^+$, where $\mathbf{U}$ is a unitary matrix, $\mathbf{U}^+$ stands for the transposed conjugate of $\mathbf{U}$. Consequently, $\mathbf{\Omega}$ can be expressed as the following convex linear combination of Hermitian, positive semidefinite matrices

$$\mathbf{\Omega} = \sum_{k=0}^{n-1} \left(P_{k+1} - P_k\right) \hat{\mathbf{\Omega}}_k \, , \tag{39.a}$$

where

$$\hat{\mathbf{\Omega}}_k \equiv \frac{\operatorname{tr}(\mathbf{\Omega})}{k+1} \left[ \mathbf{U}\, \hat{\mathbf{D}}_k \mathbf{U}^+ \right] \tag{39.b}$$

and $\hat{\mathbf{D}}_k = \operatorname{diag}(1,...,1,0,...,0)$ is the ($k$+1)-rank diagonal matrix whose $k$+1 first elements are 1 and the last $n$-($k$+1) are 0.

The expansion given by Eq. (39.a) is just the $n$D *trivial decomposition* presented in Ref [2], where it is applied and studied for the coherency matrices corresponding to 2D and 3D polarized light as well as for the coherency matrix associated with the Mueller matrix of a medium.

Moreover, the matrix $\mathbf{\Omega}$ can also be written as

$$\mathbf{\Omega} = \sum_{k=1}^{n-1} P_k \, \mathbf{\Omega}_k + \frac{\operatorname{tr}(\mathbf{\Omega})}{n} \mathbf{I}_n \, , \tag{40.a}$$

where

$$\mathbf{\Omega}_k = \frac{\operatorname{tr}(\mathbf{\Omega})}{\sqrt{n\left(k^2 + k\right)}} \left[ \mathbf{U}\, \mathbf{D}_k \mathbf{U}^+ \right], \tag{40.b}$$

$\mathbf{I}_n$ is the $n \times n$ identity matrix and $\mathbf{D}_k$ are the $n \times n$ diagonal Gell-Mann matrices.

To demonstrate Eq (40.a) it is enough to consider the relation

$$\sqrt{\frac{\left(k^2 + k\right)}{n}} \mathbf{D}_k = (k+1)\, \hat{\mathbf{D}}_{k-1} - k\, \hat{\mathbf{D}}_k \, . \tag{41}$$

### 5.3 Generalized partial entropies of a covariance matrix

The von Neumann entropy

$$S_{(n)}(\mathbf{\Omega}) = -\frac{\operatorname{tr}(\mathbf{\Omega} \ln \mathbf{\Omega})}{\operatorname{tr}(\mathbf{\Omega})} = -\frac{\sum_{i=0}^{3} \left(\lambda_i \ln \lambda_i\right)}{\operatorname{tr}(\mathbf{\Omega})} \, , \tag{42}$$

can be expressed in function of the indices of purity and, hence, they determine the value of $S_{(n)}$. In the same way as for polarization optics, respective partial entropies can be defined for each index of purity $S_{(2)}(P_i)$ $(i = 1,...,n-1)$ so that they gives entropy-like measures of the lack of purity concerning each level of detail in the purity space of the system.

## Appendix

As a basis for the expansion of a $3 \times 3$ coherency matrix **R**, we consider the following set of Hermitian, trace-orthogonal, matrices constituted by the Gell-Mann matrices plus the identity matrix

$$3\times 3 \ \textit{normalized Gell-Mann matrices}$$

$$\begin{array}{lll}
\boldsymbol{\omega}_{00}\equiv\begin{pmatrix}1&0&0\\0&1&0\\0&0&1\end{pmatrix} & \boldsymbol{\omega}_{01}\equiv\sqrt{\frac{3}{2}}\begin{pmatrix}0&1&0\\1&0&0\\0&0&0\end{pmatrix} & \boldsymbol{\omega}_{02}\equiv\sqrt{\frac{3}{2}}\begin{pmatrix}0&0&1\\0&0&0\\1&0&0\end{pmatrix}\\
\boldsymbol{\omega}_{10}\equiv\sqrt{\frac{3}{2}}\begin{pmatrix}0&-i&0\\i&0&0\\0&0&0\end{pmatrix} & \boldsymbol{\omega}_{11}\equiv\sqrt{\frac{3}{2}}\begin{pmatrix}1&0&0\\0&-1&0\\0&0&0\end{pmatrix} & \boldsymbol{\omega}_{12}\equiv\sqrt{\frac{3}{2}}\begin{pmatrix}0&0&0\\0&0&1\\0&1&0\end{pmatrix}\\
\boldsymbol{\omega}_{20}\equiv\sqrt{\frac{3}{2}}\begin{pmatrix}0&0&-i\\0&0&0\\i&0&0\end{pmatrix} & \boldsymbol{\omega}_{21}\equiv\sqrt{\frac{3}{2}}\begin{pmatrix}0&0&0\\0&0&-i\\0&i&0\end{pmatrix} & \boldsymbol{\omega}_{22}\equiv\frac{1}{\sqrt{2}}\begin{pmatrix}1&0&0\\0&1&0\\0&0&-2\end{pmatrix}
\end{array}\tag{43}$$

The notation used for these matrices is justified for the sake of simplicity as well as to emphasize the symmetry in the mathematical expressions of Gell-Mann matrices corresponding to other dimensions. In fact, leaving aside the normalization factor, the Pauli matrices are nested in the $2\times 2$ restriction of $\boldsymbol{\omega}_{01},\boldsymbol{\omega}_{10},\boldsymbol{\omega}_{11}$. Analogous nested structure exists for *n*-dimensional generalized Gell-Mann matrices.

Next, we include the explicit expressions of the $4\times 4$ Gell-Mann matrices as well as the $4\times 4$ "modified Dirac matrices" cited in the paper

$$4\times 4 \ \textit{normalized Gell-Mann matrices}$$

$$\begin{array}{llll}
\boldsymbol{\Theta}_{00}=\begin{pmatrix}1&0&0&0\\0&1&0&0\\0&0&1&0\\0&0&0&1\end{pmatrix} & \boldsymbol{\Theta}_{01}=\sqrt{2}\begin{pmatrix}0&1&0&0\\1&0&0&0\\0&0&0&0\\0&0&0&0\end{pmatrix} & \boldsymbol{\Theta}_{02}=\sqrt{2}\begin{pmatrix}0&0&1&0\\0&0&0&0\\1&0&0&0\\0&0&0&0\end{pmatrix} & \boldsymbol{\Theta}_{03}=\sqrt{2}\begin{pmatrix}0&0&0&1\\0&0&0&0\\0&0&0&0\\1&0&0&0\end{pmatrix}\\
\boldsymbol{\Theta}_{10}=\sqrt{2}\begin{pmatrix}0&-i&0&0\\i&0&0&0\\0&0&0&0\\0&0&0&0\end{pmatrix} & \boldsymbol{\Theta}_{11}=\sqrt{2}\begin{pmatrix}1&0&0&0\\0&-1&0&0\\0&0&0&0\\0&0&0&0\end{pmatrix} & \boldsymbol{\Theta}_{12}=\sqrt{2}\begin{pmatrix}0&0&0&0\\0&0&1&0\\0&1&0&0\\0&0&0&0\end{pmatrix} & \boldsymbol{\Theta}_{13}=\sqrt{2}\begin{pmatrix}0&0&0&0\\0&0&0&1\\0&0&0&0\\0&1&0&0\end{pmatrix}\\
\boldsymbol{\Theta}_{20}=\sqrt{2}\begin{pmatrix}0&0&-i&0\\0&0&0&0\\i&0&0&0\\0&0&0&0\end{pmatrix} & \boldsymbol{\Theta}_{21}=\sqrt{2}\begin{pmatrix}0&0&0&0\\0&0&-i&0\\0&i&0&0\\0&0&0&0\end{pmatrix} & \boldsymbol{\Theta}_{22}=\sqrt{\frac{2}{3}}\begin{pmatrix}1&0&0&0\\0&1&0&0\\0&0&-2&0\\0&0&0&0\end{pmatrix} & \boldsymbol{\Theta}_{23}=\sqrt{2}\begin{pmatrix}0&0&0&0\\0&0&0&0\\0&0&0&1\\0&0&1&0\end{pmatrix}\\
\boldsymbol{\Theta}_{30}=\sqrt{2}\begin{pmatrix}0&0&0&-i\\0&0&0&0\\0&0&0&0\\i&0&0&0\end{pmatrix} & \boldsymbol{\Theta}_{31}=\sqrt{2}\begin{pmatrix}0&0&0&0\\0&0&0&-i\\0&0&0&0\\0&i&0&0\end{pmatrix} & \boldsymbol{\Theta}_{32}=\sqrt{2}\begin{pmatrix}0&0&0&0\\0&0&0&0\\0&0&0&-i\\0&0&i&0\end{pmatrix} & \boldsymbol{\Theta}_{33}=\sqrt{\frac{1}{3}}\begin{pmatrix}1&0&0&0\\0&1&0&0\\0&0&1&0\\0&0&0&-3\end{pmatrix}
\end{array}\tag{44}$$

*Modified Dirac matrices*

$$\begin{array}{cccc}
\mathbf{E}_{00}=\begin{pmatrix}1&0&0&0\\0&1&0&0\\0&0&1&0\\0&0&0&1\end{pmatrix} &
\mathbf{E}_{01}=\begin{pmatrix}1&0&0&0\\0&1&0&0\\0&0&-1&0\\0&0&0&-1\end{pmatrix} &
\mathbf{E}_{02}=\begin{pmatrix}0&0&1&0\\0&0&0&1\\1&0&0&0\\0&1&0&0\end{pmatrix} &
\mathbf{E}_{03}=\begin{pmatrix}0&0&-i&0\\0&0&0&-i\\i&0&0&0\\0&i&0&0\end{pmatrix} \\
\mathbf{E}_{10}=\begin{pmatrix}1&0&0&0\\0&-1&0&0\\0&0&1&0\\0&0&0&-1\end{pmatrix} &
\mathbf{E}_{11}=\begin{pmatrix}1&0&0&0\\0&-1&0&0\\0&0&-1&0\\0&0&0&1\end{pmatrix} &
\mathbf{E}_{12}=\begin{pmatrix}0&0&1&0\\0&0&0&-1\\1&0&0&0\\0&-1&0&0\end{pmatrix} &
\mathbf{E}_{13}=\begin{pmatrix}0&0&-i&0\\0&0&0&i\\i&0&0&0\\0&-i&0&0\end{pmatrix} \\
\mathbf{E}_{20}=\begin{pmatrix}0&1&0&0\\1&0&0&0\\0&0&0&1\\0&0&1&0\end{pmatrix} &
\mathbf{E}_{21}=\begin{pmatrix}0&1&0&0\\1&0&0&0\\0&0&0&-1\\0&0&-1&0\end{pmatrix} &
\mathbf{E}_{22}=\begin{pmatrix}0&0&0&1\\0&0&1&0\\0&1&0&0\\1&0&0&0\end{pmatrix} &
\mathbf{E}_{23}=\begin{pmatrix}0&0&0&-i\\0&0&-i&0\\0&i&0&0\\i&0&0&0\end{pmatrix} \\
\mathbf{E}_{30}=\begin{pmatrix}0&-i&0&0\\i&0&0&0\\0&0&0&-i\\0&0&i&0\end{pmatrix} &
\mathbf{E}_{31}=\begin{pmatrix}0&-i&0&0\\i&0&0&0\\0&0&0&i\\0&0&-i&0\end{pmatrix} &
\mathbf{E}_{32}=\begin{pmatrix}0&0&0&-i\\0&0&i&0\\0&-i&0&0\\i&0&0&0\end{pmatrix} &
\mathbf{E}_{33}=\begin{pmatrix}0&0&0&-1\\0&0&1&0\\0&1&0&0\\-1&0&0&0\end{pmatrix}
\end{array} \qquad (45)$$